\documentclass{aastex}
\usepackage{spr-astr-addons}
\usepackage{url}
\usepackage{graphicx}

\RequirePackage{color}

\begin{document}

\title{Preliminary results for RR Lyrae stars and Classical \\Cepheids from the Vista
  Magellanic Cloud (VMC) Survey}
\slugcomment{Not to appear in Nonlearned J., 45.}
\shorttitle{First results RR Lyrae and Cepheids from VMC survey}
\shortauthors{Ripepi et al.}

\author{V. Ripepi\altaffilmark{1}}
\author{M.I. Moretti\altaffilmark{2,3}}
\author{G. Clementini\altaffilmark{3}}
\author{M. Marconi\altaffilmark{1}}
\author{M.R. Cioni$^{*}$\altaffilmark{4,5}}
\author{J.B. Marquette\altaffilmark{6}}
\author{P. Tisserand\altaffilmark{7}}

\altaffiltext{1}{INAF-Osservatorio Astronomico di Capodimonte, Via Moiariello
  16, 80131, Naples, Italy}
\altaffiltext{2}{INAF-Osservatorio Astronomico di Bologna, via Ranzani 1,
Bologna, Italy}
\altaffiltext{3}{University of Bologna, Department of Astronomy, via Ranzani 1,
40127, Bologna, Italy}
\altaffiltext{4}{University of Hertfordshire, Physics Astronomy and
  Mathematics, Hatfield AL10 9AB, UK}
\altaffiltext{5}{University Observatory Munich, Scheinerstrasse 1,
  81679 M\:unchen, Germany \\ $^{*}$~Research Fellow of the Alexander von Humboldt Foundation}
\altaffiltext{6}{UPMC-CNRS, UMR7095, Institut d'Astrophysique de Paris, F-75014, Paris, France}
\altaffiltext{7}{Research School of Astronomy \& Astrophysics, Mount Stromlo Observatory, Cotter Road, Weston ACT 2611, Australia}
\begin{abstract}
The Vista 
Magellanic Cloud (VMC, PI M.R. Cioni) survey is collecting 
$K_S$-band time series photometry of the system formed by the two Magellanic Clouds (MC) and 
the ``bridge" that connects them.  These data are used to build 
$K_S$-band light curves  of the MC RR Lyrae  stars and Classical Cepheids
and determine absolute distances and the 3D geometry  of the whole system  using the $K$-band period luminosity
($PLK_S$), the period - luminosity - color ($PLC$) and the Wesenhiet
relations applicable to these types of variables.
As an example of the survey potential we present results from the VMC
observations of two 
fields  centered respectively on the
South Ecliptic Pole and  the 30 Doradus star forming region of the Large Magellanic Cloud.  
The VMC $K_S$-band light curves
of the RR Lyrae stars in these two regions have very good photometric 
quality with typical errors for the individual data points in the range of  $\sim$ 0.02 to 0.05 mag. The Cepheids have excellent
light curves (typical errors of $\sim$ 0.01 mag). The average $K_S$ magnitudes derived for both types of variables
were used to derive $PLK_S$ relations that are in general good agreement within the errors with the
literature data, and show a smaller scatter than previous
studies.

\end{abstract}

\keywords{stars: variables: Cepheids-- stars: variables: RR Lyrae --
  galaxies: individual: LMC -- galaxies: distances and redshifts}


\section{Introduction}

The VISTA near-infrared $YJK_S$ survey of the Magellanic system
\citep[VMC; ][]{Cioni11} is an ESO public survey that is obtaining deep near infrared imaging in
the $Y$, $J$ and $K_S$  filters of a wide area across the Magellanic
system, using the VIRCAM camera \citep{Dalton_etal06} of the ESO VISTA telescope
\citep{Emerson_etal06}. \\
The main science goals of the survey are the determination of the
spatially-resolved star-formation history (SFH) and the definition of the 
three-dimensional (3D) structure of the whole Magellanic system. The VMC
observations  are devised to reach $K_S\sim 20.3$ mag, thus
allowing  us to measure sources encompassing most phases of stellar evolution:
from the main-sequence, to subgiants, upper and lower red giant branch (RGB)
stars, red clump stars, RR Lyrae and Cepheid variables, asymptotic
giant branch (AGB) stars, post-AGB stars, planetary nebulae (PNe),
supernova remnants (SNRs), etc. These different stellar populations will help us
assessing the evolution of age and metallicity within the whole MC system.\\

In this context, a very significant role is played by the radially pulsating variables, 
since these stars obey a period-mean density relation that
forms the basis of their use as standard candles to measure the distance
to the host system. The RR Lyrae stars in particular, in the $K$ band obey a
period-luminosity-metallicity ($PLZ$) relation which is
weakly affected by the evolutionary effects, the spread in stellar mass within
the instability strip, and uncertainties in the reddening corrections
\citep[see e.g.][]{long86,dallora04,coppola11}. Similarly, the Cepheid
$PL$ relation in the $K$ band is much
narrower than the corresponding optical relations, and less affected
by systematic uncertainties in the reddening and metal content 
\citep[see e.g.][]{caputo00}.

We present preliminary results 
for the  RR Lyrae stars  and Cepheids contained 
in the first two ``tiles'' completely observed by the VMC survey, namely tiles
8\_8 and 6\_6. These two tiles are centered respectively on the South Ecliptic Pole (SEP)  field and on the 
well known 30 Doradus  (30 Dor) star forming region of the Large Magellanic Cloud (LMC). 
The SEP field is particularly interesting
because this region of the sky will be continuously and repeatedly observed during the  
the commissioning phase of the astrometric Gaia satellite just after the launch in Spring 2013. 

\section{The VMC data for the variable stars}

The VMC observing strategy is described in detail in
\cite{Cioni11} to which the interested reader is referred for more information. 
Here we focus on the data acquisition for the variable stars. 

In order to obtain well sampled light curves and measure
accurate parameters (namely $K_S$ average magnitudes) for the variable sources, 
the $K_S$ observations were split into 12 epochs and distributed along
several consecutive months.  \\
We used the v1.0 VMC release ``pawprints'' (6 ``pawprints'' form a ``tile''). The pawprints were processed
by the pipeline \citep{Irwin_etal04}  of the VISTA Data Flow System \citep[VDFS,][]{Emerson_etal04} 
and retrieved from the VISTA Science Archive
\citep[VSA,][]{ham08}\footnote{http://horus.roe.ac.uk/vsa/}. \\
Usually a variable star is present in two or three not
necessarily consecutive pawprints. Hence we first calculated a weighted average
of the pawprints' $K_S$ magnitudes to obtain the ``tile''   $K_S$, which
also represents one epoch of data. 

\begin{figure}
\includegraphics[scale=.50]{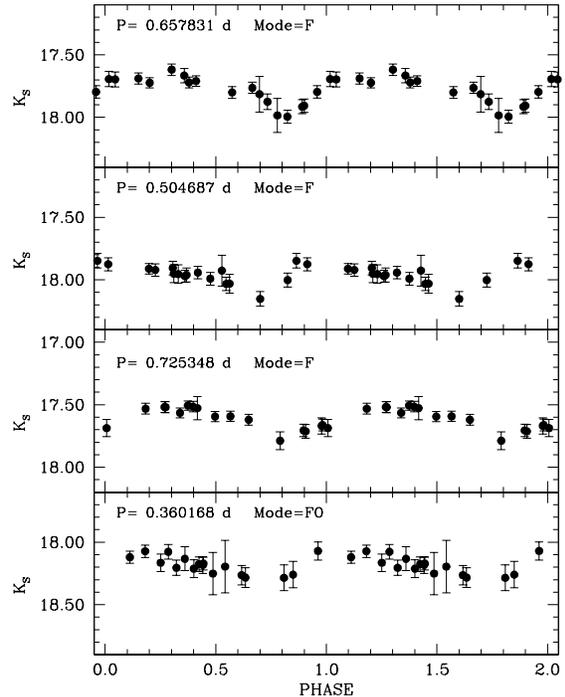}
\caption{$K_S$-band light curves of three fundamental mode (F;  3
  upper panels)  and one first overtone (FO; bottom panel) 
RR Lyrae stars in the SEP field. Internal photometric errors of the individual data points
are shown, they are typically in the range of 0.02 to 0.05
mag. Periods were taken form the EROS-2 Survey.}
\label{fig1}
\end{figure}

\begin{figure}
\includegraphics[scale=.50]{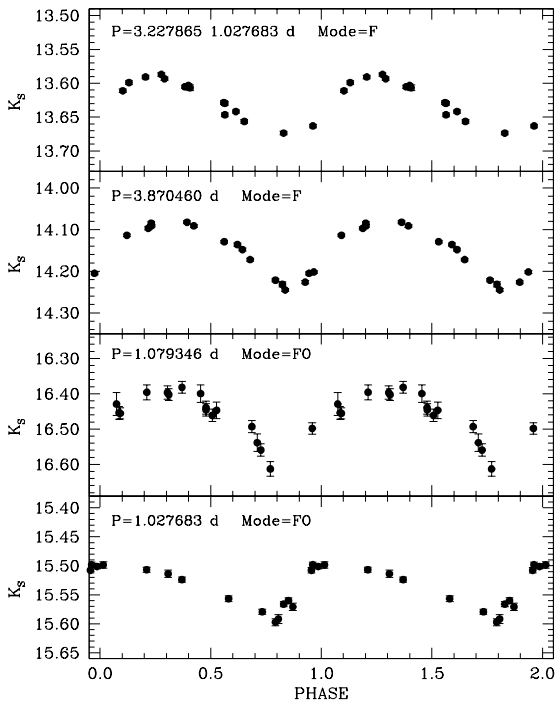}
\caption{$K_S$-band light curve for a sample of Cepheids in the SEP
  field. Note
 in the third panel the nice light curve for a faint FO Cepheid. Periods were taken form the EROS-2 Survey.}
\label{fig2}
\end{figure}

Particular care was devoted  to the 
determination of a proper Heliocentric  Julian Day (HJD) for the  $K_S$ value of  each ``tile''. 

The identification, pulsation period and epoch of maximum
light of the RR Lyrae stars and Cepheids contained in the SEP and 30 Dor regions 
are 
already available in  the photometric archives of the Microlensing surveys
OGLE and EROS-2.  

\begin{figure}
\includegraphics[scale=.50]{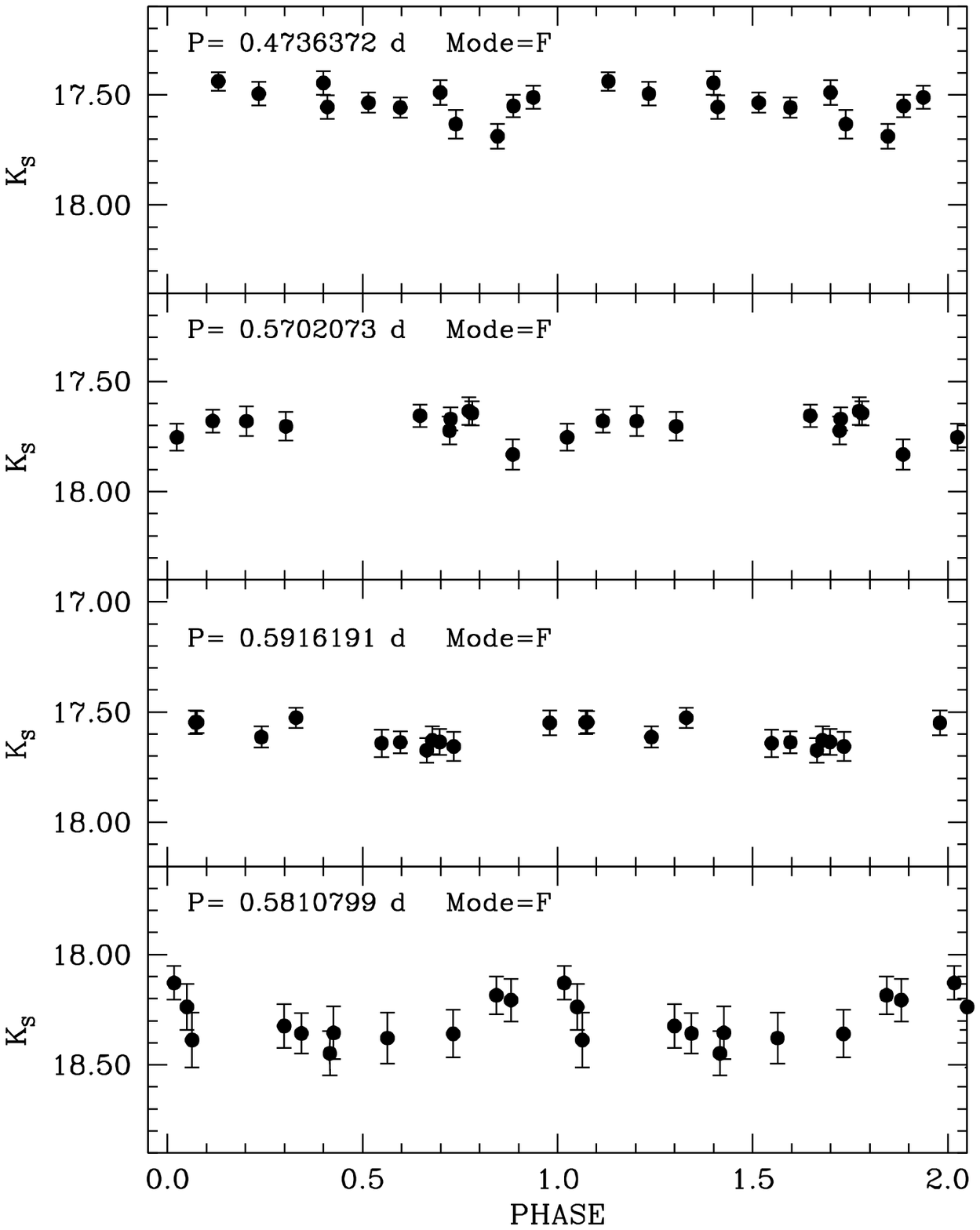}
\caption{Same as Fig.~\ref{fig1} for RR Lyrae stars in the 30 Dor
  region.  Periods were taken form the OGLE-III Survey.
}
\label{fig3}
\end{figure}
  
The OGLE surveys \citep[][and references therein]{sos08,sos09} of which stage IV is in
progress, cover an area of the MC system that extends progressively further outside
from the bar of each of the Clouds. EROS-2 \citep{tisserand07} is, at present, the
most extended of the two surveys and is the only one covering
the largest fraction of the VMC field of view including the most peripheral areas such as the SEP 
region, that in fact  only
overlaps with EROS-2.  Coordinates, periods and optical\footnote {The 
EROS-2 $blue$ channel (420-720 nm), overlaps with the $V$ and $R$ standard 
bands, and the $red$ channel (620-920 nm) roughly matches the mean 
wavelength of the Cousins $I$ band \citep{tisserand07}} 
light-curves  of the RR Lyrae and Cepheids in the Gaia SEP field were
taken from the EROS-2 catalogue and cross-matched to the VMC catalogues for the SEP tile.
The 30 Dor field is covered by both EROS-2 and OGLE-III,  
but for the present analysis we only employed the periods and optical
(Johnson-Cousins $V, I$ bands) light curves from OGLE-III
\citep[][]{sos08,sos09}. \\
As a result of the matching procedure we found 107  and 1465 
RR Lyrae variables, and 11 and 326 Classical Cepheids in the SEP
and 30 Dor fields, respectively. 

\begin{figure}
\includegraphics[scale=.50]{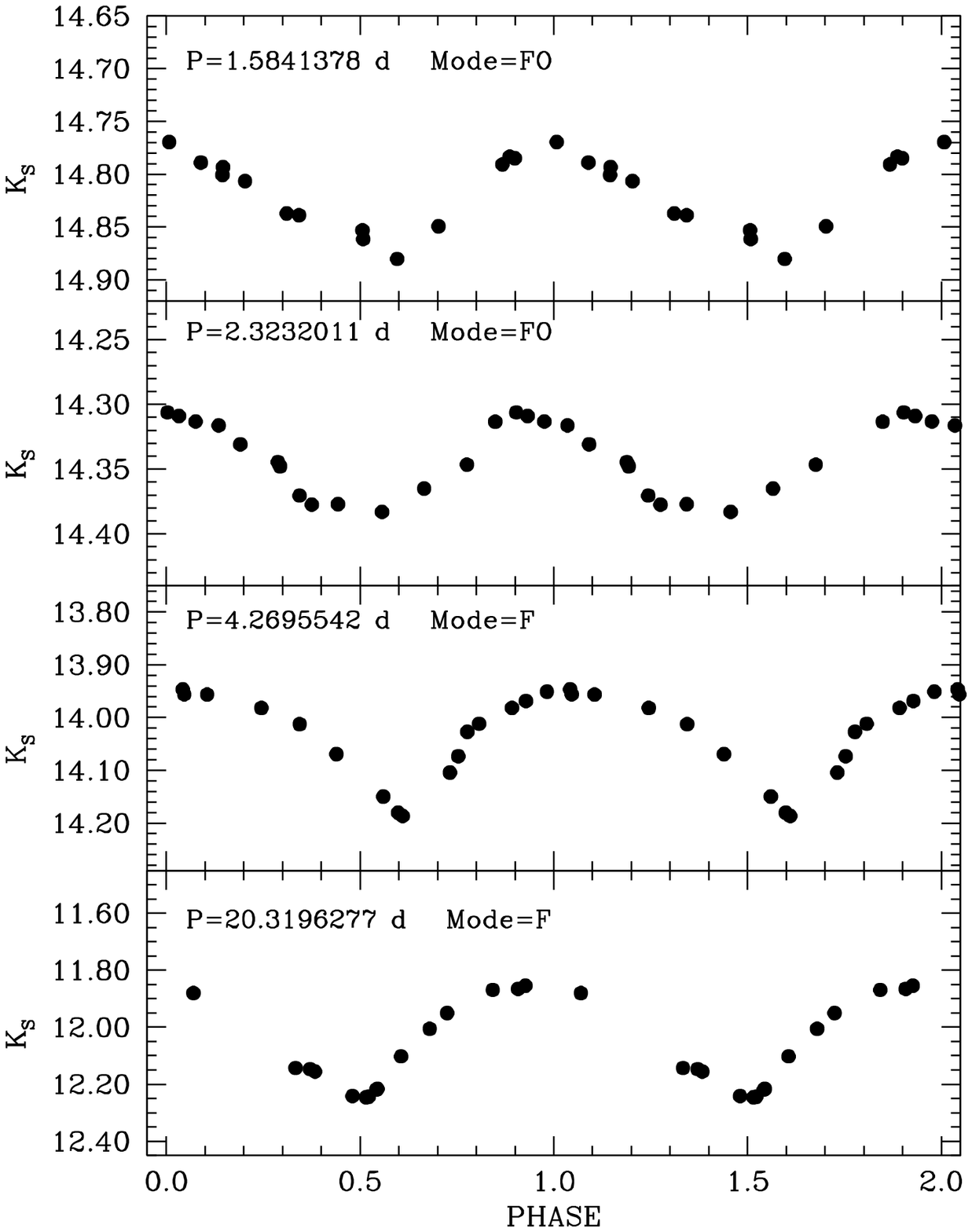}
\caption{As in Fig~\ref{fig2} but for Cepheids  in the 
30 Dor field. Typical errors of the individual data points are
of the order of 0.01-0.02 mag, hence have  same size as the data
points plotted in the figure. Periods were taken from the OGLE-III Survey. }
\label{fig4}
\end{figure}

The periods available from EROS-2 for the SEP variables and from OGLE-III for the 30 Dor variables 
were used to  fold the
$K_S$-band light curves produced by the VMC
observations and derive average $K_S$ magnitudes for the variables. 
Examples of  the VMC $K_S$-band light curves of RR Lyrae stars and Cepheids 
in the SEP and 30 Dor regions are shown in Figs.~\ref{fig1}, ~\ref{fig2}
and~\ref{fig3}, ~\ref{fig4}, respectively.  
Error bars of the individual $K_S$ measurements are shown in
the figures. The uncertainties for the Cepheids 
are of the
same size at the data-points, while they are larger for
the RR Lyrae stars, particularly in the 30 Dor region,  due to the fainter magnitudes
of these variables and the high crowding of the 30 Dor field.

\subsection{Determination of the average $K_S$ magnitudes}

In the present work we confine our discussion  to 
the RR Lyrae stars  in the SEP field and the Cepheids in the 30 Dor region only. This is because of the small number of Cepheids in the SEP tile 
(only 11), and because 
the high photometric contamination at the magnitude level of the RR Lyrae
stars in the 30 Dor field requires a more careful and detailed analysis which is still in progress.

The $K_S$-band light curves of the RR Lyrae stars are almost
sinusoidal and the amplitude of the luminosity variation in $K_S$ is also rather 
modest. However, the most accurate results for the mean magnitudes are
achieved by using the light-curve templates provided by
\cite{jones96}. Note that the use of 
templates requires accurate ephemeris and $V$-band amplitudes which, for the SEP RR Lyrae were 
calculated starting from the EROS-2 database. 

As for the Cepheids, as shown by Figs.~\ref{fig2}, ~\ref{fig4}, the light curves are well sampled and
nicely shaped. Hence, to derive the (intensity) average
$<K_S>$ we simply used a custom program written in ``c''  language that 
performed a
spline interpolation to the data.

\section{Results}

Having derived the mean $K_S$ magnitudes as discussed above, we then computed the $PLK_S$  relations for the SEP RR Lyrae stars and the 30 Dor Classical Cepheids. 

\subsection{RR Lyrae stars in the SEP field}

\begin{figure}
\includegraphics[scale=.41]{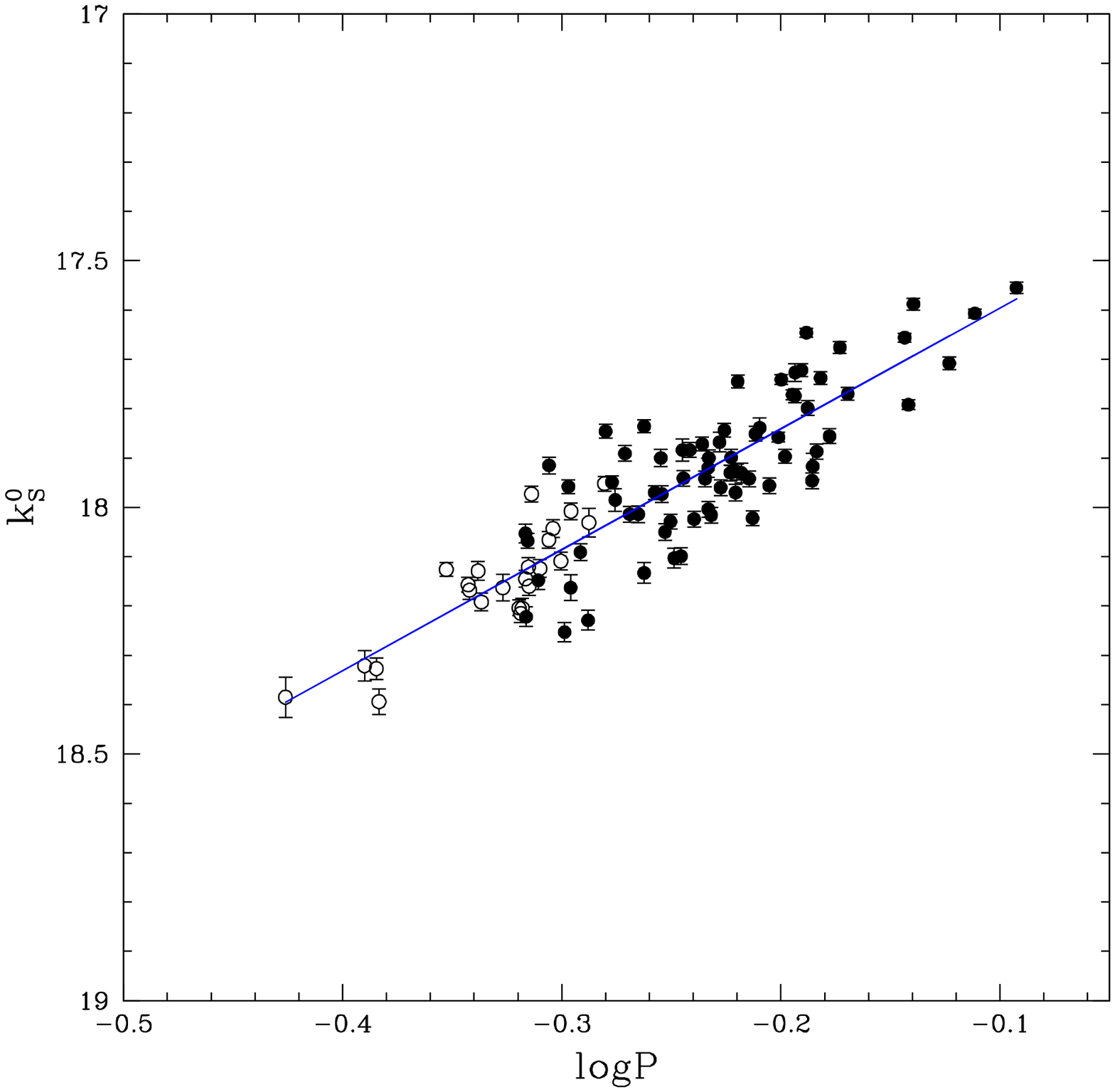}
\caption{($K_S$)-band Period-Luminosity relation for RR Lyrae variables in the
  SEP field. Open and filled circles show first overtone and fundamental mode RR Lyrae
  stars, respectively. The solid line shows the result of the least
  square fit to the data (see text for details).} 
\label{sepRR}
\end{figure}

The $PLK_S$ relation of the 95 RR Lyrae stars with good photometry in the SEP
field is shown in Fig.~\ref{sepRR}, where the periods of the FO RR Lyrae stars were
fundamentalized using the formula logP$_F$=logP$_{FO}$ + 0.127. The $K_S$
magnitudes were dereddened using the $E(B-V)$  value for the SEP field provided by
S. Rubele (private communication) and calculated as described in
\cite{rubele11}. \\
We then performed a least-square fit to the data, obtaining the
following $PLK_S$ relation:

\begin{equation}
K_S^0=(17.351\pm0.035)-(2.450\pm0.133) logP  
\label{eq1}
\end{equation}

which has an r.m.s.=0.082 mag. The higher dispersion of the SEP RR Lyrae $PLK_S$ with respect to the $PLK_S$ relations followed by the RR Lyrae in Galactic
globular clusters \citep[see, e.g.][and references
therein]{dallora04,coppola11} is likely due to a spread in distance among the 
SEP RR Lyrae along the line of sight, as well as to the possible presence of
a metallicity variance among the variables.  
We have compared our results for the SEP RR Lyrae stars with previous near-infrared
observations of RR Lyrae variables in the LMC by \cite{borissova} and \cite{szewczyk}.  Our Eq.~\ref{eq1} can be compared with Eq. 7 of
\cite{borissova} which was calculated using RR Lyrae data from
both the above mentioned papers (107 stars in total):

\begin{equation}
K_S^0=(17.47\pm0.07)-(2.11\pm0.17) logP
\label{eq2}
\end{equation}


The two equations are only marginally consistent within the errors, 
and the \cite{borissova} $PLK_S$ has a larger dispersion than ours. These occurrences can be due to both 
 geometrical (depth) effects and a metallicity spread. These aspects will be investigated in a following paper (Moretti et al. in preparation)
where we shall also provide our estimate of the LMC distance.

\subsection{Classical Cepheids in the 30 Dor Field}

\begin{table*}
\small
\caption{$K_S$-band $PL$ relations for F and FO Classical Cepheids.  The model
equation is:   $K_S^0$=$\alpha$+$\beta$ logP. The first two lines  of the table show the coefficients of the
least square fit to the data of F and FO Classical Cepheids  in the 30 Dor region presented in this paper.  The last three lines
show the coefficients of the same relations found in the
literature.}
\label{tab1}
\begin{center}
\begin{tabular}{ccccccc}
\hline
\noalign{\smallskip}
mode & $\alpha$ & $\sigma_{\alpha}$ & $\beta$ & $\sigma_{\beta}$ &
r.m.s. & source \\
\noalign{\smallskip}
\hline
\noalign{\smallskip}
F   & 16.070 & 0.017 & -3.295 & 0.018 & 0.102& This work  \\
FO & 15.580 & 0.012 & -3.471 & 0.035 &0.099 & This work \\
F   & 16.051 & 0.05   &  -3.281& 0.040& 0.108 & \citet{persson} \\
F   & 16.032 & 0.025 & -3.246& 0.036& 0.168 & \citet{groe00}  \\
FO & 15.533& 0.032 & -3.381& 0.076& 0.137 &\citet{groe00} \\
\noalign{\smallskip}
\hline
\noalign{\smallskip}
\end{tabular}
\end{center}
\end{table*}

\begin{figure}
\includegraphics[scale=.41]{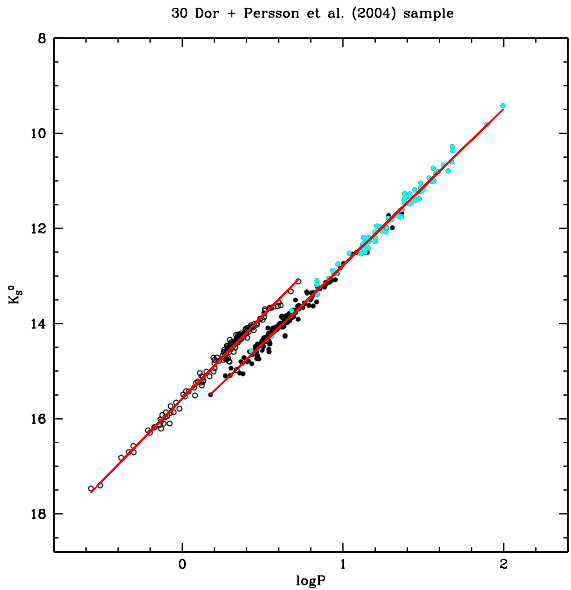}
\caption{$K_S$-band $PL$ relation for Cepheids in
  the 30 Dor field. Black open and filled circles show FO- and F-mode  
pulsators, respectively. Light blue filled circles show the F-mode
Cepheid sample by \cite{persson}. Solid lines are the result of the least
  square fit to the data (see text for details).} 
\label{figdorCep}
\end{figure}

The PLK relations of the 172 fundamental mode and the 154 first
overtone\footnote{Double mode pulsators F/FO and FO/SO were included
  in the F and FO samples, respectively.}
Classical Cepheids analysed in the 30 Dor field are shown in
Fig.~\ref{figdorCep}. Since the saturation level of the VMC survey 
in $K_S$ limits the length of the periods we are able to measure to a maximum value of about
15-20 days, we decided to complement  our data with the sample by \cite{persson}
that  includes 84 F-mode Cepheids with periods mainly ranging
between 10 and 100 days. To merge the two samples  we first transformed Persson et al.'s original 
photometry from the LCO  to the 2MASS system using the relations
by \cite{carpenter}   These data are shown as light blue circles
in Fig.~\ref{figdorCep}. To account for the variable reddening
that characterizes the 30 Dor field, we adopted the recent 
evaluations by \cite{haschke}, while for the \cite{persson}
data set we adopted the values provided by the Authors. 
Finally, we performed least-square fits to the data of  F- and
FO-modes variables separately, adopting an equation of the form 
$K_S^0$=$\alpha$+$\beta$ logP. The coefficients derived from the fits
are provided  in Table~\ref{tab1}. The same table also shows a comparison with
the literature results of \cite{persson}, who only
published the $PLK_S$ relation of fundamental-mode pulsators,  and of \cite{groe00} 
who also gives  the FO $PLK_S$ relation. An inspection of the table
reveals that 1) for the F-mode  pulsators there is general agreement
within the errors among the three studies; 2)  the present relation has 
smaller errors of the coefficients of the $PLK_S$ equation because of the 
wide range of periods spanned by the Cepheids used to derive the relation, including for the
first time in the near-infrared range a significant number of objects with
periods smaller than 5 days; 3) for the FO pulsators the agreement
with \cite{groe00} is not as good as for the F-mode Cepheids. This could
be due to the fact that the deeper photometry of the VMC survey \citep[][relies on
2MASS and DENIS data]{groe00} allowed us to reach the fainter FO-Cepheids that
populate the low-period tail of the $PLK_S$ relation.  \\

The use of the $PLK_S$ relation derived in this paper to estimate the 
distance of the LMC as well as the discussion of the 
Wesenheit and PLC relations is deferred to a forthcoming paper (Ripepi
et al. 2012, in preparation).

\section{Conclusions and future perspectives}

We have presented preliminary results on the RR Lyrae stars and the Classical
Cepheids observed by the Vista 
Magellanic Cloud (VMC) survey in two fields of the LMC, centered on the
South Ecliptic Pole (SEP) and 30 Dor region, respectively. The $K_S$ light curves
of the RR Lyrae variables have very good photometric 
quality, for the brighter Cepheids we obtained excellent
data. The average $K_S$ magnitudes derived for both kind of objects
allowed us to derive $PLK_S$ relations which were compared with the
literature, finding in general good agreement within the errors. Our
data show in any case a smaller scatter with respect to previous
studies. We expect much better results when  more data from the
VMC survey will be available. In particular we will be able to use the
$PLK_S$ relation for the RR Lyrae variables and the $PLK_S$, $PLC$ and Wesenheit
relations for Classical Cepheids to investigate both the 3D geometry and the absolute
distances to the Magellanic Cloud system.
 
\acknowledgments

M.I. Moretti thanks the Royal Astronomical Society for financial support during her
two-month stay at the  University of Hertfordshire.

\bibliographystyle{spr-mp-nameyear-cnd}

\end{document}